\documentclass[%
 reprint,
 superscriptaddress,
 amsmath,amssymb,
 aps,
]{revtex4-2}

\usepackage{xcolor}     
\usepackage{siunitx}    
\usepackage{graphicx}
\usepackage{dcolumn}
\usepackage{bm}

\newcommand{\bfSI}[2]{\text{\bfseries\SI{#1}{#2}}}

\begin{document}

\preprint{APS/123-QED}

\title{Probing Vortex Dynamics in 2D Superconductors with Scanning Quantum Microscope}

\author{Sreehari Jayaram}
\thanks{These authors contributed equally to this work.}
\affiliation{3rd Institute of Physics, University of Stuttgart, Allmandring 13, Stuttgart 70569, Germany}

\author{Malik Lenger}
\thanks{These authors contributed equally to this work.}
\affiliation{3rd Institute of Physics, University of Stuttgart, Allmandring 13, Stuttgart 70569, Germany}

\author{Dong Zhao}
\affiliation{Max Planck Institute for Solid State Research, Heisenbergstrasse 1, Stuttgart 70569, Germany}
\affiliation{Beijing National Laboratory for Condensed Matter Physics, Institute of Physics, Chinese Academy of Sciences, Beijing 100190, China}
\affiliation{School of Physical Sciences, University of Chinese Academy of Sciences, Beijing 100049, China}

\author{Lucas Pupim}
\affiliation{Institute for Theoretical Physics III, University of Stuttgart, Pfaffenwaldring 57, Stuttgart 70569, Germany}

\author{Takashi Taniguchi}
\affiliation{Research Center for Materials Nanoarchitectonics, National Institute for Materials Science, 1-1 Namiki, Tsukuba 305-0044, Japan}

\author{Kenji Watanabe}
\affiliation{Research Center for Electronic and Optical Materials, National Institute for Materials Science, 1-1 Namiki, Tsukuba 305-0044, Japan}

\author{Ruoming Peng}
\thanks{Corresponding author. E-mail: pruoming@gmail.com}
\affiliation{3rd Institute of Physics, University of Stuttgart, Allmandring 13, Stuttgart 70569, Germany}

\author{Marc Scheffler}
\affiliation{1st Institute of Physics, University of Stuttgart, Pfaffenwaldring 57, Stuttgart 70569, Germany}

\author{Rainer Stöhr}
\affiliation{3rd Institute of Physics, University of Stuttgart, Allmandring 13, Stuttgart 70569, Germany}

\author{Mathias S. Scheurer}
\affiliation{Institute for Theoretical Physics III, University of Stuttgart, Pfaffenwaldring 57, Stuttgart 70569, Germany}

\author{Jurgen Smet}
\affiliation{Max Planck Institute for Solid State Research, Heisenbergstrasse 1, Stuttgart 70569, Germany}

\author{Jörg Wrachtrup}
\affiliation{3rd Institute of Physics, University of Stuttgart, Allmandring 13, Stuttgart 70569, Germany}
\affiliation{Max Planck Institute for Solid State Research, Heisenbergstrasse 1, Stuttgart 70569, Germany}


\begin{abstract}
Magnetic dynamics at the nanoscale provide crucial insight into the behavior of superconductors. Using single-spin scanning quantum microscopy, we probe vortex dynamics in the two-dimensional superconductor NbSe$_2$. Our measurements reveal a disordered vortex glass phase that melts near the critical temperature and displays cooling-rate-dependent configurations. Surprisingly, magnetic noise persists well below $T_c$, with a strength that increases at lower temperatures---contrary to expectations. This behavior, detected via spin decoherence, points to an intrinsic origin driven by competition between supercurrent density and thermal fluctuations. Our results establish single-spin microscopy as a powerful platform for investigating fluctuations in 2D superconductors.
\end{abstract}

\maketitle


\section{Introduction}\label{intro}
Two-dimensional (2D) superconductors offer a versatile platform for exploring novel interactions and quantum phenomena. Their reduced dimensionality and consequently enhanced fluctuations manifest not only in the superfluid condensate but also in the excitations, such as vortices. In bulk type-II superconductors, vortices induced by an externally applied magnetic field form well-ordered hexagonal lattices referred to as Abrikosov vortex crystals~\cite{hess1989scanning}. In 2D superconductors, however, the behaviour of vortices is distinct. It is governed by thickness-related strong confinement, an enhanced susceptibility to disorder, as well as quantum correlation~\cite{clem1991two}. As a result, the vortices in a 2D superconductor are predicted to expand. They evolve into the so-called Pearl vortices~\cite{pearl1964current}, and arrange into a glassy state without translational invariance. Concurrently, delocalization of vortices can become pervasive~\cite{fisher1991thermal,chiang2004superconductivity,saito2016highly}, especially near the critical temperature ($T_\mathrm{c}$), where the vortex lattice is predicted to melt into a liquid phase~\cite{huberman1979melting,brandt1989thermal,koshelev1994dynamic}. The presence of defects and dislocations can also influence the vortex configuration, and a variety of vortex phases may form, such as a pinned vortex glass, a stripe vortex liquid, and ultimately, an isotropic vortex liquid~\cite{guillamon2009direct}. In disordered systems, fluctuations can drive both thermodynamic and quantum phase transitions. Examples unique to 2D systems include the Berezinskii-Kosterlitz-Thouless (BKT) transition~\cite{berezinskii1971destruction,kosterlitz1972long,ryzhov2017berezinskii}, the metal-to-insulator transition~\cite{gomory1997characterization} and many unconventional electronic phenomena~\cite{sohn2018unusual,xi2016ising,zhao2023evidence,wan2023orbital}.

To investigate nanoscale superconducting phenomena, scanning tunnelling microscopy (STM) has been widely employed to map the local density of states~\cite{hess1990vortex,roy2019melting,fischer2007scanning}. However, the tunnelling nature of such probes demands an ultra-clean conducting surface, making STM studies highly sensitive to fabrication residues. Even protective hBN cladding commonly used for 2D flakes poses challenges \cite{watanabe2004direct,dean2010boron,roy2021structure}. When approaching the 2D limit, thermal fluctuations play a crucial role for the electron behavior~\cite{fisher1991thermal,doniach1981quantum,glazman1991thermal}. They influence collective phenomena such as the dynamics of vortices. STM, as well as conventional nanoscale probes~\cite{de1999scanning}, such as magnetic force microscopy (MFM), and nano superconducting quantum interference devices (nano-SQUIDs), do not simultaneously offer the required spatial and temporal resolution to capture this emergent magnetic dynamics. 

To overcome these limitations and visualize local magnetic excitations with the spatiotemporal resolution necessary for the 2D superconductors, scanning quantum microscope (SQM) based on nanoscale magnetometry with single nitrogen vacancy (NV) centers in diamond, has been developed~\cite{balasubramanian2008nanoscale,maletinsky2012robust}. Single vortices and the effective penetration depth were characterized in a quantitative manner~\cite{thiel2016quantitative,pelliccione2016scanned}. The quantum sensing community has also extended the toolbox for SQM with numerous pulse sequences to extend its dynamic range~\cite{degen2017quantum,levine2019principles}. Among others, screening effects associated with superconductivity were investigated, demonstrating the ability of SQM to detect surface charge noise in the diamond substrate~\cite{monge2023spin}. So far, SQM has primarily been applied to address the static response of high-Tc superconductors. Extending its application to two-dimensional superconductors holds great promise, since the local dynamic response of 2D superconductors should be particularly rich, but remains largely unexplored, which is the purpose of this work. 

We visualize the local magnetic response of few-layer NbSe$_2$ with SQM. In contrast to the hexagonal Abrikosov vortex lattice observed in bulk superconductors, we detect a distorted vortex glass. By varying the cooling rate, we capture the stochastic process of vortex glass condensation. Furthermore, dynamic spin decoherence measurements reveal an unusual temperature dependence of the fluctuations with enhanced magnetic noise persisting to temperatures well below the critical temperature. By combining high temporal and spatial resolution within our SQM technique, we have succeeded in recording the in-plane magnetic dynamics within the disordered vortex glass that forms in a 2D superconductor due to enhanced thermal fluctuations and incomplete screening within the superconductor. 

\section{Local vortex excitation in the 2D superconductor}\label{sec_excitation}

\begin{figure}
    \includegraphics[width=\columnwidth]{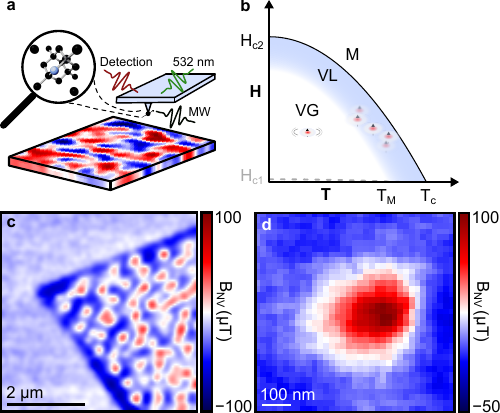}
    \caption{\textbf{Scanning quantum microscopy of the vortex state in 2D superconductors.} (a) Schematic of the scanning quantum microscope used to probe local magnetic responses at the nanoscale. (b) Phase diagram of a \SI{5.1}{\nano\metre} thick NbSe$_2$ flake as a function of temperature and magnetic field, identifying three distinct phases: vortex glass, vortex liquid, and a metallic state. (c) Magnetic field map of the vortex glass phase, showing pronounced field contrast and highlighting the sample boundary. (d) High-resolution scan resolving a single vortex in the NbSe$_2$ flake.
    }
    \label{setup}
\end{figure}

Our measurements were performed on samples out of thin layers of the 2D superconductor 2H-NbSe$_2$ using a \SI{1.8}{\kelvin} cryogenic atomic force microscope (AFM) as illustrated in Fig. \ref{setup}a. As the NV center approaches the NbSe$_2$ sample (S1), the spin of a single NV can detect the local magnetic responses of the NbSe$_2$ irrespective of the top hBN cladding (see supplementary information (SI) section 1 for details). The NbSe$_2$ flake consists of two regions with different thicknesses: \SI{5.1}{\nano\metre} (7 layers) and \SI{12.6}{\nano\metre}. An optical image and AFM scan are shown in Fig. S1 in the SI. In the thinner NbSe$_2$, three distinct phases — vortex glass (VG), vortex liquid (VL), and metallic (M) state — are illustrated, as shown in Fig. \ref{setup}b. These phases have been previously characterized through transport measurements, revealing the transitions between superconducting and dissipative states~\cite{zhang2020nonreciprocal}. The melting temperature is denoted as $T_\mathrm{M}$, while$H_\mathrm{c1}$ and $H_\mathrm{c2}$ represent the lower and upper critical fields, respectively, for a type-II superconductor. Fig. \ref{setup}c presents a scan recorded for the vortex glass state with a diamagnetic response along the flake edge at a temperature of \SI{2}{\kelvin}, indicative of the Meissner effect. Near \SI{10}{\kelvin} ($>T_c$), any magnetic response has vanished (Fig. S4 in SI), confirming that the observed magnetic behaviour at \SI{2}{\kelvin} originates from the superconducting state. Since the flake's thickness is significantly smaller than the penetration depth $\lambda$ of bulk NbSe$_2$ ($\lambda\sim$\SI{125}{\nano\metre})~\cite{callaghan2005field,fletcher2007penetration}, the external magnetic field is only partially screened by the 2D superconductors. The effective field screening is approximately \SI{-100}{\micro\tesla}. 

Here, we also observe a positive magnetic signal, attributed to superconducting vortices that persist throughout almost the entire superconducting phase. Notably, vortices emerge under external fields as low as a few mT, significantly below the reported first critical field ($B_{c1}$) for bulk NbSe$_2$. Due to the few-layer thickness of the sample, the magnetic field readily penetrates the material, locally suppressing superconductivity and forming vortices at a much lower external field. The vertical confinement in the very thin NbSe$_2$ layer causes an expansion of the vortex, following the Pearl model \cite{pearl1964current}. From a high-resolution scan of a single vortex in Fig. 1d, the size of the vortex is expanding compared to the vortex in the bulk, matching the expected scaling behaviour for Pearl vortices in thin-layer superconductors \cite{fridman2025anomalous}.

Sample characterization with Atomic Force Microscopy (AFM) (Fig. S1 in SI) reveals significant residue on the flake surface after sample fabrication. Such a dirty top surface poses technical challenges for other local probes such as scanning tunnelling microscopy (STM) \cite{zomer2014fast,pizzocchero2016hot}. The SQM technique put forward here is obviously more robust. It enables high-resolution and non-invasive mapping of the superconducting vortices. The surface residues may introduce small deviations in the measured magnetic field map, but these can be corrected a posteriori by correlating field maps with AFM topography data.

\section{Arrangement of the vortices in 2D superconductor}\label{sec_conf}
\begin{figure}
    \centering
    \includegraphics[width=\columnwidth]{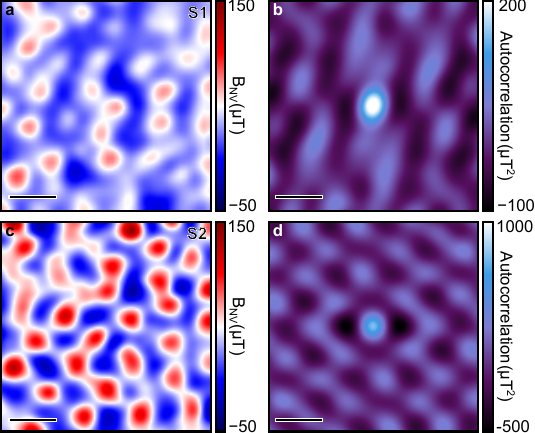}
    \caption{\textbf{Vortex arrangements in 2D NbSe\boldmath$_2$ with varying thickness.} (a, b) Magnetic field map and corresponding autocorrelation of a thin (\SI{5.1}{\nano\metre}) NbSe$_2$ sample S1 on an oxide substrate and encapsulated with hBN. The vortex configuration appears disordered, showing weak spatial correlations and broad, smeared autocorrelation peaks. (c, d) Magnetic field map and autocorrelation of a thicker (\SI{11.59}{\nano\metre}) NbSe$_2$ sample S2, fully encapsulated with hBN, revealing enhanced vortex hexagonal ordering. All scale bars are \SI{1}{\micro\metre}.}
    \label{conf1}
\end{figure}

In the thin NbSe$_2$ sample (S1), substrate roughness, disorder and finite-size effects play an important role in the superconducting response. Due to these local effects, vortices can become distorted and elongated, so their shape deviates from the regular vortex shape typically observed in a pristine superconductor. Because of the multiband nature of superconductivity in NbSe$_2$~\cite{zehetmayer2010experimental,noat2015quasiparticle}, multiple-flux-quanta vortices~\cite{gozlinski2023band} may appear. Fig. \ref{conf1}a shows an example data set where no clear hexagonal vortex arrangement can be discerned. The autocorrelation pattern in Fig. \ref{conf1}b reveals a highly distorted vortex arrangement with elongated and irregular features that blur the vortices of the correlation peaks. An underlying hexagonal pattern can barely be recognized.  As also seen in Fig. S5 of the SI, the vortex configuration is strongly constrained by the sample boundary, and features in the autocorrelation signal align with the orientation of the edge. In contrast, hexagonal ordering appears more clearly toward the center, but it remains significantly influenced by local disorder. 

Thin samples also show features that appear like elongated vortices,  as indicated in Fig. \ref{conf1}a, for temperature well below $T_\mathrm{c}$. We attribute these line-shaped magnetic field structures to basically circular vortices that rapidly fluctuate between distinct positions in space, i.e. they are somewhat delocalized, effectively causing the elongated shapes in the SQM images. This vortex motion is likely mediated by the underlying disorder and the enhanced thermal fluctuations in a 2D superconductor~\cite{schmid1973theory}. For the sake of comparison, we performed similar measurements on a larger and thicker NbSe$_2$ sample ($>$\SI{10}{\nano\metre}, S2) encapsulated on both sides with hBN to ensure flat and pristine samples (Fig. S2 in SI). In this case, vortices arrange in a more orderly fashion, and a hexagonal vortex lattice is observed in most of the sample area (Fig. \ref{conf1}c). Subsequent autocorrelation analysis of different regions (Fig. \ref{conf1}d) reveals a natural Abrikosov vortex arrangement, although it remains distorted due to the competition with geometric confinement and spatial disorder. We note that the lack of knowledge about the specific and non-hexagonal vortex arrangement in thinner samples may be important for the interpretation of, for instance, transport experiments. This is frequently overlooked, yet may be important for a proper deduction of the symmetry of the superconducting ground state from such experimental data.
    
\section{Melting transition of the vortex lattice}\label{sec_melting}

Previous transport studies on thin NbSe$_2$ samples have also provided indirect evidence that the melting of the vortex glass takes place more easily than in bulk systems. The strong thermal fluctuations can depin vortices and drive the melting transition~\cite{zhang2020nonreciprocal}. The observation of non-zero resistivity across a broader temperature window, but still below $T_\mathrm{c}$, indeed suggested that vortices detach from their pinning site and make a transition into the vortex liquid phase. Such melting can be further promoted by the presence of dislocations~\cite{soibel2000imaging,guillamon2009direct}, therefore, the vorex liquid evolves differently as the cooling speed is varied.

\begin{figure}
    \centering
    \includegraphics[width=\columnwidth]{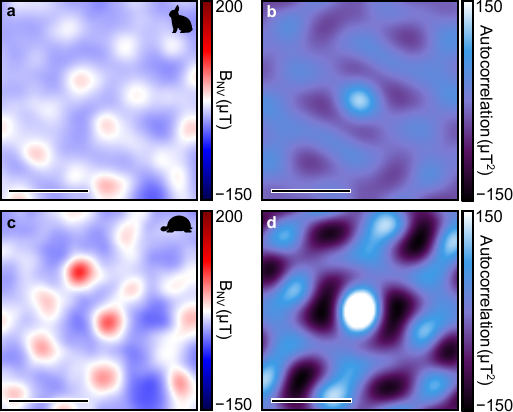}
    \caption{\textbf{Cooling rate–dependent vortex arrangements in \bfSI{5.1}{\nano\metre} NbSe\boldmath$_2$ sample S1.} (a) SQM stray field map of a thin NbSe$_2$ flake after rapid cooling (from \SI{6.3}{\kelvin} to \SI{2}{\kelvin} within 1 minute), showing weak magnetic contrast. (b) Corresponding autocorrelation reveals a disordered vortex configuration with poor spatial correlation. (c) Magnetic field map after slow cooling (from \SI{6.3}{\kelvin} to \SI{2}{\kelvin} over several hours), showing enhanced vortex contrast. (d) Autocorrelation of (c) indicates more distinct local vortex ordering and stronger magnetic response. All scale bars are \SI{1}{\micro\metre}.
    }
    \label{coolingrate}
\end{figure}

We have performed such a study with varied cooling speeds. For each cooling sequence, the sample S1 is “initialized“ by raising the temperature to \SI{6.3}{\kelvin}. This is still below $T_\mathrm{c}$=\SI{6.8}{\kelvin}, but at this temperature only a weak magnetic field response remains, and isolated vortices are no longer visible in a large-area scan (See Fig. S6 in SI). The zoomed-in scan reveals weak magnetic signals without distinct vortex features (See Fig. S7 in SI). Subsequently, the sample is cooled to base temperature at the selected rate. This procedure is repeated for different cooling rates from 1 minute up to 10 hours until the base temperature of \SI{2}{\kelvin}. The SQM measurements for the lowest and highest cooling rates are shown in Fig. \ref{coolingrate}a and \ref{coolingrate}c. Although both measurements focus on the same location of the few-layer sample, the supercurrent distribution for fast and for slow cooling speed shows a different Meissner field response as well as vortex intensity.

The difference is further enhanced in the autocorrelation graph, shown in Fig. \ref{coolingrate}b and \ref{coolingrate}d. We observe an enhancement in absolute autocorrelation intensities under slow cooling conditions. Distinct variations in autocorrelation patterns suggest a redistribution of supercurrent densities driven by different cooling protocols. For a rapid cool down from the melted state, one may anticipate that vortices are not localized or pinned as strongly as upon slow cooling. The abrupt temperature drop causes immediate pinning at surrounding disorder sites, and these pinned vortices remain more susceptible to thermal fluctuations, resulting in a more ‘volatile’ vortex configuration with vortex delocalized on a small length scale and an overall weaker magnetic response. In contrast, slow cooling from the melted state should allow thermal fluctuations to compete with disorder pinning, so that a more stable vortex lattice is gradually formed.

\section{Dynamics in the vortex state}
\begin{figure}
    \includegraphics[width=\columnwidth]{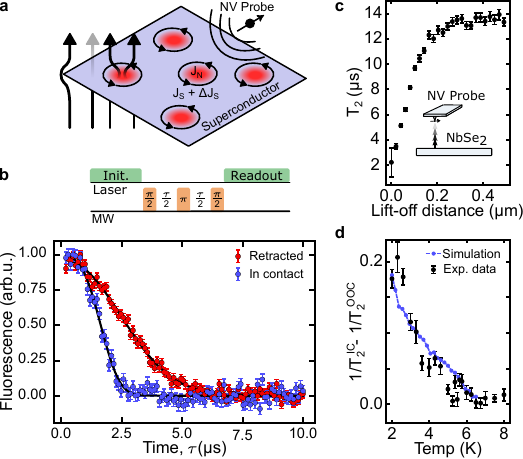}
    \caption{\textbf{Unexpected magnetic noise in 2D superconductors.} (a) Schematic of a local NV probe detecting supercurrent-induced magnetic fluctuations in a 2D superconductor. Magnetic fields are only partially screened in the thin superconducting layer. (b) Spin Hahn echo $T_2$ measurements with the NV probe in contact with and retracted from the sample, revealing enhanced dephasing near the surface. Inset: Spin Hahn echo pulse sequence (see SI section 2.4). (c) Extracted $T_2$ times as a function of lift-off distance. These data suggest that there is a critical length scale for magnetic fluctuations $>$\SI{100}{\nano\metre}. (d) Enhanced decoherence rate between in-contact and out-of-contact (tip-sample distance greater than \SI{1}{\micro\metre}) conditions (1/$T_{2}^{IC}$-1/$T_{2}^{OOC}$), measured below the superconducting transition temperature (black dots). This rate reveals magnetic noise originating from the superconducting sample and exhibits an inverse temperature dependence, in good agreement with numerical simulations (blue dots).
    }
    \label{ac1}
\end{figure}

Beyond static measurements, the NV spin offers dynamic magnetic field sensing capabilities across a broad frequency range~\cite{levine2019principles}. By employing tailored pulse sequences, we can isolate and target specific noise sources affecting the spin, thereby improving the spectral selectivity of dynamic measurements. The pulse capability makes NV-based techniques essential tools for investigating fluctuation phenomena in 2D superconductors from DC to GHz. Notably, significant fluctuations are expected to arise both from the thermal excitation of quasi-particles, even at temperatures below $T_\mathrm{c}$~\cite{chatterjee2022single,dolgirev2022characterizing}, and from fluctuations of the superconducting order parameter~\cite{CurtisXYModel}. As illustrated in Fig. \ref{ac1}a, the supercurrent fluctuation can induce strong transverse magnetic noise, leading to the dephasing of the NV electron spin.

We probe the underlying magnetic noise arising from NbSe$_2$ by conducting Hahn echo measurements of the NV electron spin in proximity to the sample. Compared to the out-of-contact condition (tip-sample distance greater than \SI{1}{\micro\metre}), there is a noticeable reduction in the $T_2$ time, as can be seen in Fig. \ref{ac1}b.  The Hahn echo $T_2$ measurements indicate the presence of magnetic noise in the vortex glass state with frequencies spanning from tens of \si{\kilo\hertz} to \si{\mega\hertz}. We also characterize spin decoherence as a function of the lift-off distance from the sample surface of the tip and find a clear trend, as shown in Fig. \ref{ac1}c: spin coherence time decreases as the probe approaches the superconducting sample. From this behaviour, we deduce a characteristic length scale of several hundred nanometers of the noise source (i.e. vortex movement) underneath the tip. It also represents direct evidence that the vortex in a 2D superconductor itself acts as the primary source of decoherence. Since no significant decoherence is observed in other thick superconducting samples, such as the \SI{300}{\nano\metre} cuprate samples shown in Fig. S9, we attribute the noise to the intrinsic two-dimensional nature of the thin NbSe$_2$ layer. To quantify the magnetic noise, the spin decoherence rate, $1/T_2$, was measured at different temperatures and compared with the spin decoherence rate in the out-of-contact condition as shown in Fig. \ref{ac1}d. This data set shows that the NbSe$_2$-induced change in the decoherence rate is larger at lower temperatures and much less pronounced when the temperature approaches or is above $T_\mathrm{c}$. 

Since fluctuations of the superconducting order parameter are expected to be largest in the vicinity of $T_\mathrm{c}$~\cite{CurtisXYModel} and contributions from quasi-particle excitations should decay with temperature~\cite{chatterjee2022single,dolgirev2022characterizing}, this observation seems surprising at first sight. However, it is important to note that $1/T_2$, unlike $1/T_1$, is primarily sensitive to frequencies in the MHz range such that the aforementioned contributions may not play a dominant role. Instead, the thermally activated motion of the vortex glass is likely the primary factor limiting $T_2$. While vortex motion itself decreases as temperature is lowered, it is the resulting magnetic field fluctuations at the NV site that drive decoherence. The magnetic field distribution around vortices in thin superconductors depends critically on the ratio of penetration depth $\lambda$ and thickness $d$. Since $\lambda$ diverges at $T_c$ and decreases as temperature is reduced, the local magnetic field at the NV center becomes increasingly sensitive to vortex position, leading to enhanced decoherence. To understand this more explicitly, we solved a Langevin equation, as a simple model for vortex motion at finite temperature, and then computed the resulting magnetic field fluctuations at the NV center to obtain a temperature-dependent decoherence rate (see SI section 3 for details). Even this simple model, requiring minimal assumptions, reproduces the temperature-dependent behaviour of  $1/T_2$ observed in Fig. \ref{ac1}d (model shown as a solid line). See Supplemental Material [url will be inserted by publisher] for additional theoretical derivations and further data analysis, which includes Refs.~\cite{Tetienne2012,kossler2012magnetic,prozorov2006magnetic,Holmlund1996,Cywinski2008}.

\section{Conclusion}
{In conclusion, we demonstrate nanoscale spatiotemporal imaging of vortices in few-layer 2D superconductors, enabling direct visualization of vortex phenomena previously accessible only indirectly. In non-uniform NbSe$_2$, we observe dynamic, deformed vortices and direct evidence of vortex melting, where the configuration is highly cooling-rate dependent, reflecting the interplay of thermal fluctuations and pinning. Both the spatial arrangement and dynamic behaviour of vortices at low magnetic fields have significant implications for the collective macroscopic response of the system, as measured by transport, nuclear magnetic resonance (NMR), etc. Our technique also reveals unexpected magnetic noise from vortex fluctuations well below $T_c$, indicating persistent low-temperature dynamics and a glassy vortex phase. These findings establish SQM as a powerful probe of low-energy fluctuations and dynamic states in unconventional superconductors.}
~\cite{ge2024charge,tanaka2001soliton,kirtley1996direct,iguchi2023superconducting}. 

\vspace{36pt}
\section*{Acknowledgements}
S.J., M.L., R.P., R.S. J.W. acknowledge funding support by the DFG via FOR 2724 and the RTG GRK 2642, the EU via the AMADEUS project, BMBF via project QSOLID 13N16159 and QMAT 03ZU2110HA, and the Zeiss Foundation through the QPhoton. J.H.S. acknowledges funding support from the DFG Priority Program SPP 2244. L.P.~and M.S.S.~acknowledge funding by the European Union (ERC-2021-STG, Project 101040651---SuperCorr). Views and opinions expressed are however those of the authors only and do not necessarily reflect those of the European Union or the European Research Council Executive Agency. Neither the European Union nor the granting authority can be held responsible for them.


\newpage

\bibliography{apssamp}

\end{document}